\documentclass{kluwer}    % Specifies the document style.
\usepackage{epsfig}
\def\msun{\mbox{M$_{\odot}$}}
\def\mvir{\mbox{M$_{\rm v}$}}
\def\ms{\mbox{M$_{\rm s}$}}
\def\md{\mbox{M$_{\rm d}$}}
\def\vm{\mbox{V$_{\rm max}$}}
\def\fd{\mbox{f$_{\rm d}$}}

\def\fg{\mbox{f$_{\rm g}$}}
\def\rs{\mbox{r$_{\rm s}$}}
\def\infrate{\mbox{$\dot{\Sigma}_{\rm g}$}}
\def\Sg{\mbox{$\Sigma_{\rm g}$}}

\def\sfr{\mbox{$\dot{\Sigma}_{\rm s}$}}
\def\ssd{\mbox{$\Sigma_{\rm s}$}}
\def\ssg{\mbox{$\Sigma_{\rm g}$}}
\def\kms{\mbox{km s$^{-1}$}}

\begin{document}                                                                                   
\begin{article}
\begin{opening}         
\title{Star formation history in the solar neighborhood: the link between stars
and cosmology\thanks{Invited talk}}
\author{V. \surname{Avila-Reese}, C. \surname{Firmani}\footnote{Also 
Osservatorio Astronomico di Brera, Italy} and X. \surname{Hernandez}}  
\runningauthor{V. Avila-Reese, C. Firmani and X. Hernandez}
\runningtitle{Solar neighborhood star formation history}
\institute{Instituto de Astronom\'{\i}a, U.N.A.M., A.P. 70-264, 04510 M\'exico, D.F.}
%\date{April 15, 1993}

\begin{abstract}
Using a cosmological galactic evolutionary approach to model the Milky Way,
we calculate the star formation history (SFH) of the solar neighborhood. The 
good agreement we obtain with the observational inferences suggests
that our physical model describes accurately the long term/large spatial trends   
of the local and global Milky Way SFH. In this model, star formation 
is triggered by disk gravitational instabilities and self-regulated by an 
energy balance in the ISM. The drivers of the SFH are the cosmological gas 
infall rate and the gas surface density determined by the primordial spin 
parameter. A $\Lambda$CDM cosmology was used throughout.

\end{abstract}

\keywords{cosmology: dark matter --- Galaxy: evolution --- Galaxy: halo --- 
galaxies: formation --- solar neighborhood --- stars: formation}

\end{opening}

\section{Introduction}

Galaxies are the ``ecosystems'' where stars are born and evolve 
interacting with the ISM. On the other hand galaxies are the structural 
units of the universe as a whole. Thus, the study of galaxy formation 
and evolution connects
in a natural way stellar and ISM astrophysics with cosmology. The Milky Way 
(MW) is among the best studied galaxies, in particular at the solar neighborhood.
The available observations for the solar neighborhood provide valuable 
information not only about its present-day properties but also about its past 
history. For example, the possibility of resolving stars and constructing a 
color-magnitude diagram (CMD) for the solar neighborhood allows us in principle 
to recover its star formation history (SFH). 

 Star formation (SF) is a key ingredient for galaxy formation and evolution.
Most models of galaxy formation in a cosmological context extend the 
hierarchical assembly of cold dark matter (CDM) halos to luminous galaxies 
using {\it ad hoc} and/or empirical recipes for calculating the SF (e.g., White
\& Frenk 1991; Kauffmann, White, \& Guiderdoni 1993; Baugh, Cole, \& Frenk 1996).
An important improvement in this kind of approaches is to link the SF process
to the structure, dynamics and hydrodynamics of modeled galaxies, and to test
the model with detailed observations, for example the SFH in the solar 
neighborhood. 

We have developed an approach of disk galaxy formation and evolution in a 
cosmological context which includes a physically self-consistent
scheme to calculate the SF process of the growing disk
(Firmani \& Avila-Reese 2000; Avila-Reese \& Firmani 2000, hereafter FA00 and 
AF00 respectively). Using this approach and in the light of the 
available observational information, we will explore the drivers of the local
and global SFH in the MW and their potential connection to the cosmological 
background.

\section{Observational inferences}

An objective inference of the SFH in the solar neighborhood comes from
by comparing synthetic CMDs to the observed one (e.g., Chiosi et al. 1989). 
These CMD inversion methods have been improved more recently with a more 
rigorous statistical analysis (Tolstoy \& Saha 1996; Hern\'andez, Valls-Gabaud 
\& Gilmore 2000 and more references therein). For example, the advanced 
Bayesian analysis introduced by Hern\'andez et al., allows the  
recovery of the underlying SFH without the need of assuming
any {\it a priori} SFH. With the {\it Hipparcos} 
catalog the time resolution of this technique is $\sim 0.05$ Gyr, however, 
the age range which it allows to explore is small (the last 3 Gyr); this limitation
is related to the incompleteness of the catalog. With a standard parametric 
maximization technique, Bertelli \& Nasi (2001) were able to recover the SFH 
along the whole life of the disk solar neighborhood, at expenses of a low time 
resolution. Their method is useful to describe
only the {\it general trend of the SFH over the total life of the 
system}, and not its detailed shape. The robust conclusion we can extract
from this study is that the average SF rate (SFR) over the period 10-6 Gyr (in 
look back time) could not have been larger than that over the 6-0 Gyr period.
In Fig. 1a, we plot the surface SFR \sfr\ vs. look back time 
corresponding to the 
solution for the model with more degrees of freedom of those used in 
Bertelli \& Nasi (jointed circles). That the solar neighborhood SFR in the 
past could not have been very different from the present-day one 
is also suggested by the birthrate parameter $b$ and the differential 
metallicity distribution of G-dwarfs (e.g., Boissier \& Prantzos 1999). 

A quantity more accurately determined than the SFH shape, is the look 
back time at 
which stars began to form in the solar neighborhood. Using the photometric
and kinematic data of the {\it Hipparcos} satellite, Binney, Dehnen, 
\& Bertelli (2000) inferred an age for the disk at the solar radius of 
$11.2\pm 0.75$ Gyr (horizontal shaded box in Fig. 1a). Regarding its present-day
SFR, observational inferences give values around $3-5 \ \msun$Gyr$^{-1}$pc$^{-2}$
(vertical shaded box).

Methods alternative to the CMD inversion have been also used to infer the 
SFH in the solar neighborhood, among them the construction of the 
stellar age distribution using the empirical correlation of chromospheric 
activity in G-K dwarfs with their age (e.g., Rocha-Pinto et al. 2000; see 
references on other methods therein). These methods are subject of strong
observational uncertainties and biases. For example, the chromospheric 
activity-age relation presents a strong scatter of around a factor 2 in age 
which should be included as a time smoothing kernel on the inferred SFH. 
Interestingly, if such a smoothing is applied on the results of Rocha-Pinto
et al. (2000), a trend similar to that reported in Bertelli \& Nasi (2001) 
appears.

\section{A theoretical approach}

As it was discussed above, the observational data in the solar neighborhood
allow to infer a general trend of its SFH, which due to stellar diffusion it may
actually correspond to the average SFH of an annulus of $\sim 2-3$ kpc.
We will present predictions regarding local and global SFHs for the
MW modeled using a deductive disk galaxy evolution approach
(FA00; AF00). This approach 
is based on the hierarchical formation scenario, assuming that disks form 
by gas accretion more than by repeated merging of sub-unities which would destroy 
the disk. A brief description of our approach is as follows: 

A disk forms {\it inside-out} within a growing CDM halo. The evolution of 
the halo is set by its mass aggregation history (MAH).
For a given present-day total virial mass \mvir, a statistical set of 
MAHs is calculated from the initial Gaussian density fluctuation field once 
the cosmology and power spectrum are fixed (Avila-Reese, Firmani \& 
Hern\'andez 1998). Given the MAH, the halo 
density profile is calculated with a generalization of the
secondary infall model, based on spherical symmetry and adiabatic 
invariance, but allowing for non-radial orbits 
%with fixed pericenter-to-apocenter ratios $\epsilon$ 
(Avila-Reese et al. 1998). The halo density profile depends on the MAH. 
%and $\epsilon$. For $\epsilon \approx 0.2$, 
The average MAH yields density profiles
very close to the Navarro, Frenk \& White (1997) profile, while 
the other less probable MAHs yield a variety of density profiles which 
also agree with what is found in N-body cosmological 
simulations (Avila-Reese et al. 1998, 1999). 
A fraction \fd\ of the mass of each collapsing spherical shell is 
transferred in a virialization time into a central disk gas layer.
Assuming angular momentum conservation, a given gas element of 
the shell falls into the equatorial plane at the position where 
it reaches centrifugal equilibrium. The specific angular momentum of 
each shell is obtained from the derivative of $\lambda$, assumed
to be constant in time. We calculate
the halo contraction due to the gravitational drag of the infalling gas. 
A nearly exponential disk surface density arises naturally in this 
inside-out scheme of disk formation. The $\lambda$ parameter determines 
the scale length and surface density of the disk. 
%According to analytical and numerical studies, $\lambda$ has a 
%log-normal distribution with an average of $\sim0.05$ and a 
%width in the logarithm of $0.5-1.0$.

\noindent {\bf Star formation}. SF at a given radius in the gaseous disk is 
triggered by gravitational instabilities, i.e. whenever the Toomre parameter 
$Q_{g} (r)$
%=v_g(r)\kappa(r)/\Sg(r) 
is less than a given threshold, $q$
%$v_g$ and $\Sg$ are the gas turbulent rms velocity and surface density, 
%and $\kappa$ is the epicyclic frequency; 
(numerical and observational studies suggest $q\approx 2$).
The SF rate (SFR) is calculated
from the equation that relates the energy input rate (mainly due to SNae) 
to the (turbulent) energy dissipation rate assuming that $Q_g (r)$ is always 
equal to $q$ at all radii, i.e. we allow only for the stationary solution 
of a {\it self-regulated} SF mechanism (Firmani \& Tutukov 1994; Firmani,
Hernandez, \& Gallagher 1996). A key parameter in this scheme is the turbulent 
dissipation time $t_d(r)$ which we approximate as in Firmani et al. (1996).
Estimates of $t_d(r)$ for the solar neighbourhood obtained in 
compressible magneto-hydrodynamic simulations of the ISM (Avila-Reese 
\& V\'azquez-Semadeni 2001) are in agreement with the Firmani et al. 
approximation. The evolution of the stellar populations is followed with
a parametrization of simple population synthesis models. A Salpeter IMF 
with minimal and maximal masses of 0.1\msun\ and 100\msun, and solar 
metallicities were used. The azimuthally averaged dynamics of the 
evolving gas and stellar disks coupled with the dark halo are treated 
by solving the corresponding hydrodynamical equations. 
%With our method we obtain the temporal evolution of the gas infall rate on the 
%disc $\infrate (r,t)$, the SFR $\sfr(r,t)$, the disc gas and star surface 
%densities, $\Sg(r,t)$ and $\ssd(r,t)$, and the circular velocities due to the 
%disc and halo, V$_{\rm d}(r,t)$ and V$_{\rm h}(r,t)$, respectively at all radii. 
%In our model the halo structure and the disc gas infall rate are directly 
%related to the cosmological background.

\begin{table}
\caption[]{Properties of the MW: observations and model results}
%label{tbl-1}
\begin{tabular}{lcr} 
\hline 
Observable \ \ \ \ \ \ \ \ \ \ \ \ \ \ \ \ \ \ \ \ \ \ \ \ \ \ \ &  Predicted value $^{a}$ \ \ \ \ \ \ \ \ \ \ \ \ & Observed value\\ 
\hline
$V_c(50)$ $^{b}$ \ [\kms] & 208 & $ 206 \pm 10$ \\
\rs \ [kpc] & 3.0  & 3.0 $\pm$ 0.5 \\
M$_{\rm s}$ $^{c}$ \ [$10^{10}$ \msun] & 4.4 & 4-5\\ \hline
$\mvir$  \ [$10^{12}$ \msun] &  2.8  & 1-4\\
$\vm^{d}$ \ [\kms] & 235  & 220 $\pm$ 10  \\
$L_B$ \ [$10^{10} L_{B_{\odot}}$] & 1.7 & 1.8 $\pm$ 0.3\\ 
r$_{B}$ $^{e}$ \ [kpc] & 4.3 & 4-5 \\
$\Sigma_{\rm 0,K}$ $^{f}$ \ [$L_{K\odot} pc^{-2}$] & 810 & $1000 \pm 200$\\   
\fg         & 0.23 & $0.15-0.20$ \\
SFR \ [\msun yr$^{-1}$] & 2.9 & 2-6 \\ \hline
{\it Solar neighborhood} \\ \hline
\ssd  \ [\msun pc$^{-2}$] & 41.3 & 43 $\pm$ 5 $^{g}$   \\
\ssg  \ [\msun pc$^{-2}$] & 13.6  & 13 $\pm$ 3  \\
\sfr  \ [\msun Gyr$^{-1}$ pc$^{-2}$] & 3.1 & 3-5 \\ 
$B-K$ \ [mag] & 3.15 & 3.13 \\ \hline
\end{tabular}

\begin{flushleft}
$^{(a)}$ {The MW model was obtained tuning three input parameters in order
to reproduce the first three quantities, which are constraints and not
predictions.}
$^{(b)}$ {Circular (asymptotic) velocity at 50 kpc.}
$^{(c)}$ {Stellar (disc+bulge) mass.}
$^{(d)}$ {This maximum rotation velocity does not take into account
the nuclear and bulge region and it is for Galactocentric radii smaller than 15 kpc.}
$^{(e)}$ {$B-$band disc scale length.}
$^{(f)}$ {Disc central surface brightness in the $K-$band, taking a mass to light 
ratio of 1.0 for this band.}
$^{(g)}$ {This estimate includes the contribution of stellar remnants.}

\end{flushleft}
\end{table}

\vspace{0.35cm}

\noindent {\bf Modeling the Milky Way.} 
For a given cosmology (here: $\Omega_{m}=0.3$, $\Omega_{\Lambda}=0.7$,
$H_{0}=65$ \kms Mpc$^{-1}$, $\sigma_{8}=0.9$) a galaxy model is 
completely determined by the virial mass \mvir, the MAH, the spin 
parameter $\lambda$, and the disk mass fraction \fd. These factors 
and their statistical distribution are related to 
the cosmological background, and we have shown that they are able
to produce disk galaxies which match the Tully-Fisher relation and 
its scater, and several correlation across the Hubble sequence (FA00; AF00). 
In order to calculate a model representative of the MW, we have to chose 
correctly these input factors. Fortunately, each one of them is tightly related to 
a present-day MW feature, although with some (small) degeneracy. 
In Hernandez, Avila-Reese \& Firmani (2001, HAF01) we showed that the sensitivity
of the local SFH to this degeneracy and to the uncertainty of the observables
that constrict the input parameters are small.

The MAH determines mainly the galaxy color index. 
Unfortunately the MAH cannot be described by one simple parameter 
and the color index of the MW is not well determined. Therefore, does not make
sense to fix a very particular MAH for the MW. We shall use the most
probable realization of the MAHs (the average one), but we also will explore
other possible MAHs.     
As a matter of fact, the $B-K$ color estimated for the MW disk ($\sim
3.3$, see Kent et al. 1991 for the $K-$magnitude estimate) is similar
to the average color of galaxies of its type, suggesting   
that the MW is indeed an average galaxy.

The total mass \mvir, the spin parameter $\lambda$, and the disk mass fraction
\md\ are fixed in such a way that the modeled MW obeys the observable constraints
related to these parameters: the circular velocity at 50 kpc, the disk scale
lenght \rs, and the disk+bulge mass, mainly the mass contained in stars, \ms. For 
the average MAH, these constraints fix $\mvir=2.8 \ 10^{12}
\msun$, $\lambda=0.02$, and \fd=0.021 (the fiducial model). For this model 
the constraints and predictions, and the corresponding observational estimates 
are given in Table 1 (see HAF01 for the references). The shortcoming of the 
model is an excesively peaked rotation curve w.r.t. observations. This
is associated with the large central concentrations of CDM halos. The predicted 
SFH does not change significantly if a shallow core is introduced in the 
MW CDM halo (HAF01). 

\vspace{0.35cm}

\noindent {\bf Predicted local and global star formation histories.}
In Fig. 1a we show the SF and gas infall histories, \sfr\ and \infrate, of 
the fiducial model at the radius $R_0=8.5$ kpc. There is a remarkable 
agreement with the 
observational estimates. The disk at $R_0$ begins to form at look back time
$\sim 11$ Gyr ($z\approx 2$), the SFR attains a maximum at $8-6$ Gyr  
($z\approx .9-.6$) and then it decreases by a factor of $\sim 2$ towards the 
present. With these SF and gas infall histories the present-day stellar and gas 
surface densities are 41.3 and 13.8 \msun pc$^{-2}$, respectively.
%The corresponding evolution of the stellar and gas surface densities are 
%shown in Fig. 1b.

In HAF01 we have experimented with a wide range of MAHs (the MAH influences
stonghly the SFH), and we have found that the SFH predicted with the most 
probable (average) MAH is the best in matching the observational data. 
Using Padova stellar evolutionary models we computed synthetic CMDs for 
this SFH taking a solar metallicity for the last 2 Gyr, and one third
the solar before, as a first approximation to the enrichment history. The
obtained $B-K$ color index is 3.15, in excellent agreement with estimates
for the solar neighborhood, $B-K=3.13$ (Binney \& Merrifield 1999). In the
future we shall compare directly the {\it Hipparcos} CMD with our theoretical
CMD (thanks to A. Bressan for suggesting this to us). Repeating the experiment
for SFHs resulting from other MAHs, we obtain $B-K$ colors up to 0.25 mag redder 
and 0.15 mag bluer than 3.15. Again, the average MAH seems to be the optimal choice 
for the Galaxy. 

The global SF and gas infall histories (in \msun\ yr$^{-1}$) for the fiducial 
MW model are shown in Fig. 1b. The maximum SFR is 
attained at $z\approx 1.3$, then
decreases by a factor of 1.5 towards $z=0$. A nearly constant SFH is a common
feature in the evolution of our modeled disk galaxy population.
Our predicted SF and gas infall histories do not deviate from the 
constraints of the phenomenological chemo-spectrophotometric models 
(e.g., Boissier \& Prantzos 1999). Therefore, we can expect 
our model to be also in agreement with the data regarding 
metallicities and abundance gradients in the solar neighborhood and the Galaxy. 

\begin{figure} % figure 1
\vspace{5pc}
{\epsfig{file=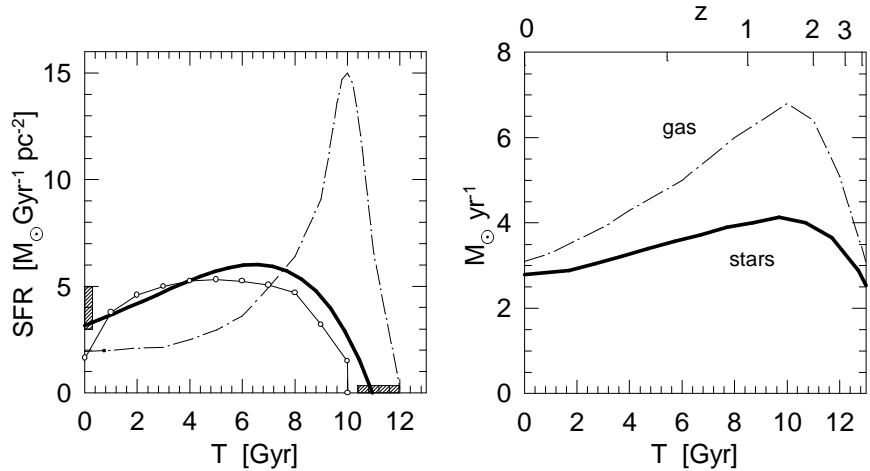, width=0.47\hsize}}
\caption[]{{\it (a) Left:} Surface SF and gas accretion histories for the 
solar neighbourhood corresponding to the fiducial model (thick solid
and dot-dashed lines). The jointed circles give the inferences 
of Bertelli \& Nasi 2001, and the boxes are independent observational
estimates for the age of the solar neighborhood and its present-day
SFR. {\it (b) Rigth:} Global SF and gas accretion histories for the fiducial
MW model.}
\end{figure}

\section{Drivers of the SFH and concluding remarks}

Within our SF scheme, the two important drivers of the SFH are
the gas infall history and the gas disk surface density.
The former is related to the cosmological MAH while the latter 
depends mainly on the spin parameter
$\lambda$. For a roughly $1-\sigma$ range in the MAHs and $\lambda$'s, 
most of the predicted global SFHs of disk galaxies present  
a shallow maximum at $z\approx 1-2.5$ and then slightly decrease
until $z=0$ by a factor $<3$ (see Fig. 1b, for the case
of a MW-like galaxy). The shape of the SFH 
of the universe inferred from high-redshift observations does not agree
with this nearly constant behaviour of the SFH of disk galaxies,
suggesting that other galaxy populations (starburst dwarfs?, bulges?, 
ellipticals?) dominated at larger redshifts in the integral SFR of the universe. 

The local SFH (e.g. at radius $R_0$) also is controled by 
the (cosmological) gas infall history and the local gas surface density. The 
stationary self-regulation mechanism combined with the disk gravitational
instability triggering criterion lead to a $\sfr (r)\propto \Sg^n$ law, 
with $n\approx2$ in rough agreement with observations. In the light of 
our inside-out scenario, the formation epoch $T$ of the MW disk at a given 
radius (e.g. $R_0$) depends on the angular momentum of the infalling gas, i.e. 
on $\lambda$. On the other hand, $\lambda$ determines 
\rs. Interestingly, once we fixed $\lambda$ to reproduce the observed \rs, 
the predicted $T_0$ is in agreement with the observational estimates,
suggesting that $T_0$ and \rs\ could indeed be related to the only parameter
$\lambda$.
Note that we have used $H_0=65$ \kms Mpc$^{-1}$; cosmic time is inversely 
proportional to $H_0$, so that the predicted $T_0$ also depends on $H_0$. A 
similar connection, although more trivial, appears to be
related to the $B-K$ color index of the solar neighborhood and its SFH. 
According to our model, both of them are related to the MAH, and for
the average MAH the predictions agree with the observational
inferences.

The overall self-consistency of our cosmological MW model predictions and 
the agreement found with the inferred SFHs in the solar
neighborhood, suggest that the main ingredients of the SF process in the
MW were correctly taken into account. Our stationary solution may 
represent the long term/large spatial trends of the SFH in the solar
neighborhood. Complementary mechanisms of SF or SF enhancement could 
introduce fluctuations of short temporal and spatial character. 

\acknowledgements
V.A. and X.H. thank the Organizing Committee for providing them finantial support in 
order to assist to this conference. This work was partially supported by CONACyT grant
J33776-E to V.A.

\end{article}

\begin{thebibliography}{}

\bibitem{}Avila-Reese V., \& Firmani C., 2000, Rev. Mex. Astron.
Astrofis., 36, 23 (AF00) 
\bibitem{}Avila-Reese V., \& V\'azquez-Semadeni E., 2000, ApJ, in press 
(astro-ph/0101397)
\bibitem{}Avila-Reese V., Firmani C., \& Hern\'{a}ndez X., 1998, ApJ, 505,
37 
%\bibitem{}Avila-Reese V., Firmani C., Klypin A., Kravtsov A., 1999, MNRAS, 310, 527 
\bibitem{}Baugh C.M., Cole S., Frenk C.S., 1996, MNRAS, 283, 1361
\bibitem{}Bertelli G., \& Nasi E., 2001, AJ, 121, 1013.
\bibitem{}Binney J., Dehnen W., \& Bertelli G. 2000, MNRAS, 318, 658
\bibitem{}Binney J., Merrifield M. 1998, ``Galactic Astronomy''
(Princeton Univ. Press), 556
\bibitem{}Boissier S., \& Prantzos N., 1999, MNRAS, 307, 857 
\bibitem{}Chiosi C., Bertelli G., Meylan G., Ortolani S., 1989, A\&A, 219,
167
\bibitem{}Firmani C., \& Avila-Reese V. 2000, MNRAS, 315, 457 (FA00)
\bibitem{}Firmani C., \& Tutukov A.V. 1994, A\&A, 288, 713
\bibitem{}Firmani C., Hern\'{a}ndez, X., \& Gallagher, J. 1996, A\&A, 308,
403
\bibitem{}Hern\'andez X., Avila-Reese V., \& Firmani C., 2001,
astro-ph/0105092 (HAF01)
\bibitem{}Hern\'andez X., Valls-Gabaud D., \& Gilmore G. 2000, MNRAS,
316, 605
\bibitem{}Kauffmann G., White, S.D.M., \& Guiderdoni, B. 1993, MNRAS, 264,
201
\bibitem{} Kent S., Dame T., Fazio G., 1991, ApJ, 378, 131
\bibitem{}Navarro J., Frenk C.S., White S.D.M., 1997, ApJ, 490, 493
\item Rocha-Pinto, H., Scalo, J., Maciel, W.J., \& Flynn, C. 2000a,
ApJ, L115
%\bibitem{}Rocha-Pinto H., Maciel W.J., Scalo J., \& Flynn, C. 2000b, A\&A, 358, 869
\bibitem{}Tolstoy E., Saha A., 1996, ApJ, 462, 672.
\bibitem{}White S.D.M, \& Frenk C.S. 1991, ApJ, 379, 52

\end{thebibliography}
\end{document}